\documentclass[trackchanges,twocolumn]{aastex701}

\usepackage{amsmath}
\usepackage{multirow}
\usepackage{adjustbox}
\usepackage{graphicx}
\usepackage{comment}
\usepackage{float}
\usepackage{placeins}
\begin{document}

\title{Particle Acceleration and Depolarization in the Protostellar jet knots HH~80 and HH~81}%

\author[0000-0001-9716-5319]{A.G. Cheriyan}
%\altaffiliation{Corresponding author: A. G. Cheriyan}
\email{email@gmail.com}
\affiliation{Department of Earth and Space Sciences \\
Indian Institute of Space Science and Technology \\
 Thiruvananthapuram 695547, India}

\author[0000-0002-3477-6021]{S. Vig}
\email{email@gmail.com}
\affiliation{Department of Earth and Space Sciences \\
Indian Institute of Space Science and Technology \\
 Thiruvananthapuram 695547, India}

%\collaboration{20}{(AAS Journals Data Editors)}

\author[0000-0001-9829-7727]{Nirupam Roy}
\email{email@gmail.com}
\affiliation{Department of Physics \\
Indian Institute of Science \\
Bangalore 560012, India}
\affiliation{Department of Physics \\ New Mexico Institute of Mining and Technology \\ Socorro, NM 87801, USA \\}

\author[0000-0001-9371-7104]{Samir Mandal}
\email{email@gmail.com}
\affiliation{Department of Earth and Space Sciences \\
Indian Institute of Space Science and Technology \\
 Thiruvananthapuram 695547, India}

\author[0000-0003-2862-5363]{C. Carrasco-Gonz\'{a}lez}
\email{email@gmail.com}
\affiliation{Instituto de Radiastronom\'{i}a y Astrof\'{i}sica (IRyA)\\
Universidad Nacional Aut\'{o}noma de M\'{e}xico (UNAM)\\
Morelia, Michoac\'{a}n, M\'{e}xico}

\author[0000-0002-4731-4934]{A. Rodr\'{i}guez-Kamenetzky}
\email{email@gmail.com}
\affiliation{Instituto de Astronom\'{i}a Te\'{o}rica y Experimental (IATE)\\
Universidad Nacional de C\'{o}rdoba (UNC)\\
C\'{o}rdoba, Argentina}

\author[0000-0003-1933-4636]{A. Pasetto}
\email{email@gmail.com}
\affiliation{Instituto de Radiastronom\'{i}a y Astrof\'{i}sica (IRyA)\\
Universidad Nacional Aut\'{o}noma de M\'{e}xico (UNAM)\\
Morelia, Michoac\'{a}n, M\'{e}xico}

%\collaboration{all}{The Terra Mater collaboration}

%% Use the \collaboration command to identify collaborations. This command
%% takes an optional argument that is either a number or the word "all"
%% which tells the compiler how many of the authors above the command to
%% show. For example "\collaboration[all]{(DELVE Collaboration)}" wil include
%% all the authors above this command.
%%
%% Mark off the abstract in the ``abstract'' environment. 
\begin{abstract}
Linearly polarized emission is a powerful tracer of magnetic field geometry and particle acceleration in protostellar jets. We present a polarimetric study of the HH objects HH~80 and HH~81 from where non-thermal emission has been confirmed through spectral index measurements at low frequencies. We carried out observations of HH~80 and HH~81 with the Karl G. Jansky Very Large Array in 4-6~GHz. Unlike the inner jet knots, no linear polarization is detected towards the knots HH~80 and HH~81. We place a $3\sigma$ upper limit of $30~\mu$Jy on the polarization intensity, corresponding to fractional polarization limits of $\Pi_{\max}\approx0.02$ and $0.01$ for HH~80 and HH~81, respectively. To interpret this non-detection, we assess the conditions for synchrotron polarization and the impact of depolarization mechanisms. The shock cooling parameter $\chi_\mathrm{s}$ is lower in these outermost HH objects than in the inner knots, indicating that the reverse shocks in HH~80-81 are less efficient at accelerating relativistic electrons compared with the inner knots. Moreover, Faraday depolarization appears severe: the dispersion in the estimated rotation measure $\sigma_{\rm RM}\sim400~\mathrm{rad~m^{-2}}$ is comparable to or larger than observed RM values themselves. This is consistent with strong fluctuations and turbulence. Together with beam depolarization, these effects can suppress the observable fractional polarization flux densities below the detectable thresholds. We conclude that reduced acceleration efficiency (when compared to inner knots) and strong depolarization account for the absence of polarized emission towards HH~80 and HH~81. 

\end{abstract}

%% Keywords should appear after the \end{abstract} command. 
%% The AAS Journals now uses Unified Astronomy Thesaurus (UAT) concepts:
%% https://astrothesaurus.org
%% You will be asked to selected these concepts during the submission process
%% but this old "keyword" functionality is maintained in case authors want
%% to include these concepts in their preprints.
%%
%% You can use the \uat command to link your UAT concepts back its source.

%% From the front matter, we move on to the body of the paper.
%% Sections are demarcated by \section and \subsection, respectively.
%% Observe the use of the LaTeX \label
%% command after the \subsection to give a symbolic KEY to the
%% subsection for cross-referencing in a \ref command.
%% You can use LaTeX's \ref and \label commands to keep track of
%% cross-references to sections, equations, tables, and figures.
%% That way, if you change the order of any elements, LaTeX will
%% automatically renumber them.

\section{Introduction} \label{sec:Introduction}

Protostellar jets are among the most energetic manifestations of early stellar evolution during the early phases of young stellar objects (YSOs). They play a key role in removing excess angular momentum and magnetic flux from the system, thereby enabling continued accretion of material from the surrounding envelope onto the protostar \citep{2018A&ARv..26....3A}. The accretion process is mediated by an accretion disk, which channels gas and dust inward, while the associated bipolar jets carry away angular momentum along the rotational axis. The launching of these jets is thought to occur in the immediate vicinity of the protostar, where gas from the inner accretion disk is lifted and centrifugally accelerated along large-scale helical magnetic field lines \citep{1994ApJ...429..781S,1994ApJ...429..797S}. This magneto-centrifugal acceleration mechanism, originally proposed for jets from accretion disks around black holes \citep{1976MNRAS.176..465B,1976Natur.262..649L}, has been adapted to explain protostellar jets \citep{1983ApJ...274..677P, 1988ApJ...328L..19S}. The subsequent collimation of the wide-angle outflow into a narrow jet is thought to occur at larger distances from the protostar, predominantly through magnetic processes. Magnetic field lines anchored in the rotating accretion disk are predicted to spiral into a helical configuration, producing toroidal components that exert hoop stresses \citep{1994ApJ...429..781S,2019FrASS...6...54P,2025ApJ...978L..31R}. Direct measurements of magnetic fields in protostellar jets remain extremely challenging due to limited spatial resolution and the need for complex modeling, often requiring various assumptions. Nonetheless, as magnetic fields are a fundamental ingredient in both the launching and collimation of protostellar jets \citep{2010HiA....15..438C,2014ApJ...792..116Z}, there have been several attempts to measure them through various methods \citep{2010Sci...330.1209C,2017A&A...597A..43G,2018NatCo...9.4636L,2019FrASS...6...66C}.

An effective probe for simultaneously determining the structure and strength of magnetic fields in astrophysical jets is the observation of polarized emission \citep{2010Sci...330.1209C,2025ApJ...978L..31R}. When electrons travel at relativistic velocities in the presence of a magnetic field, they emit linearly polarized synchrotron radiation. In the centimetre-wavelengths, synchrotron spectra typically display negative spectral indices \citep{1970ranp.book.....P}. This method has been highly successful in studying the magnetic fields of jets from active galactic nuclei (AGNs) \citep{1997ARA&A..35..607Z,2017Galax...5...11G,2021ApJ...923L...5P,2022Galax..10..102P}. In contrast, protostellar jets exhibiting much lower velocities, of the order of $200$--$1000~\mathrm{km~s^{-1}}$ \citep{1994ApJ...436..125H,2001ARA&A..39..403R}, rendering the detection of relativistic electrons more challenging. In these systems, the radio continuum is generally dominated by thermal free-free emission. This thermal process is characterized by a positive spectral index in radio at low frequencies and an absence of intrinsic polarization \citep{1986ApJ...304..713R}. Nonetheless, a small number of protostellar jets associated with both low- and high-mass YSOs have shown radio emission with negative spectral indices \citep{1989ApJ...346L..85R,2017ApJ...851...16R,2017ApJ...840...36O,2018MNRAS.474.3808V,2019MNRAS.486.3664O,2019A&A...623A..77S,2023MNRAS.525.2172C}, consistent with non-thermal synchrotron radiation from relativistic electrons. These electrons are likely to be accelerated in regions of strong shocks formed where the fast thermal jet collides with the ambient medium \citep{2015A&A...582L..13P,2020SSRv..216...29P,2016ApJ...818...27R,2017ApJ...851...16R}. In these shock fronts, diffusive shock acceleration could accelerate particles to relativistic energies \citep{1949PhRv...75.1169F}. Definitive confirmation of synchrotron radiation in protostellar jets requires the detection of linearly polarized radio emission. Since polarized emission constitutes only a fraction of the total continuum flux, and protostellar jets are generally faint in radio, such measurements demand extremely sensitive observations. Although linear polarization (LP) has been detected towards the inner lobes of protostellar jets \citep{2010Sci...330.1209C,2025ApJ...978L..31R}, measurements towards the terminal lobes with non-thermal emission, located at much larger distances from the central protostar, have not been reported. In this study, we investigate the presence of LP in the outer lobes of the prominent jet system HH~80-81.

The HH~80-81 jet is one of the largest known, highly collimated, and most luminous protostellar jet in our Galaxy \citep{1989ApJ...346L..85R, 1993ApJ...416..208M, 2023JApA...44...57M}. It is powered by the massive protostar IRAS~18162-2048 (hereafter I18162), with properties consistent with a zero-age main-sequence (ZAMS) B0 spectral type. Located at a distance of approximately $1.4~\mathrm{kpc}$ \citep{2020ApJ...888...41A}, I18162 resides in a complex star-forming environment where millimeter-wavelength observations have revealed 25 dense cores. Among these, the most massive core, MM1, has been identified as the driving source of the HH~80-81 jet \citep{2019A&A...623L...8B}. The distance between the driving source and the jet knots HH~80-81 is $\sim~2.1$~pc. Synchrotron emission has been detected in the inner lobes (S~15 and S~13)($\sim$0.4~pc) of the jet via measurements of LP \citep{2010Sci...330.1209C}, indicating a helical magnetic field configuration \citep{2025ApJ...978L..31R}.

The morphology and kinematics of the HH~80-81 jet have been examined extensively at optical wavelengths by \citet{1998AJ....116.1940H}, who reported knot velocities in the range of $600$--$700~\mathrm{km\,s^{-1}}$.
X-ray observations detect hot, shock-generated plasma in both HH~80 and HH~81, with inferred shock velocities reaching $320~\mathrm{km~s^{-1}}$ \citep{2004ApJ...605..259P}. Radio continuum studies reveal that HH~80 and HH~81 have negative spectral indices ( $\alpha \approx -0.2$ to $-0.6$), confirming that the emission is dominated by non-thermal synchrotron emission \citep{1995ApJ...449..184M,2018MNRAS.474.3808V,2023JApA...44...57M}. The combination of a large-scale, magnetized, and massive jet, with multi-epoch and multi-frequency radio observations, source's relative proximity and presence of non-thermal emission make HH~80 and HH~81 ideal targets for investigating linear polarization from knots at large distances from protostars. In order to examine the polarization properties toward HH~80 and HH~81, we have conducted full Stokes radio observations using the Karl G. Jansky Very Large Array (VLA) between 4 and 6 GHz, where the observations were carried out as two separate runs. The data from both runs were concatenated to generate images in Stokes parameters (I, Q, and U), enabling a comprehensive analysis of LP.

The structure of this paper is as follows: Section~\ref{sec:Observations} describes the observational data and reduction process. In Section~\ref{sec:Results}, we present our results, which are interpreted and discussed in Section~\ref{sec:Discussion}. Finally, a summary of our conclusions is provided in Section~\ref{sec:Conclusion}.

%%%%%%%%%% Fig 1 begins %%%%%%%%%%%%%
\begin{figure*}
    \centering
    \includegraphics[width=\linewidth]{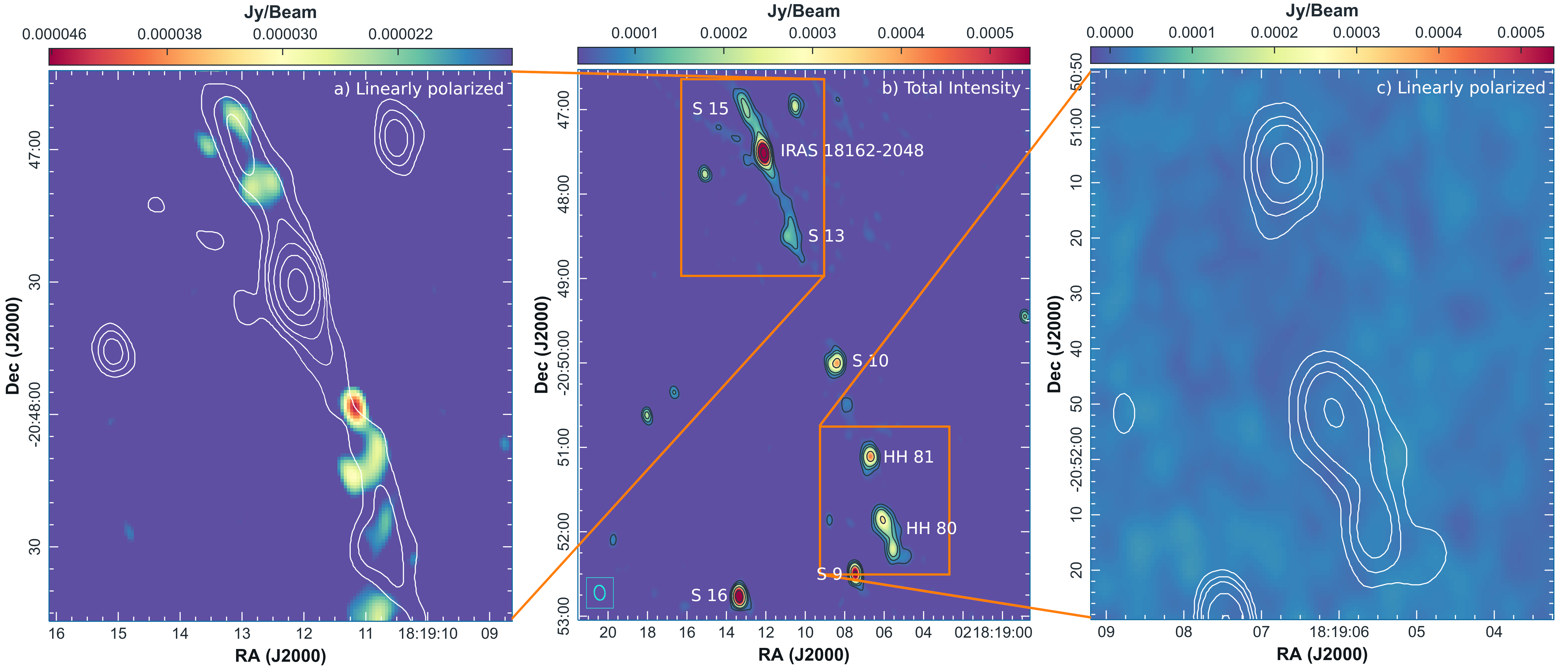}
    \caption{ (a) Linearly polarized emission observed toward the inner jet lobes of I18162. The white contours overlaid on the image are the Stokes I contours ($(6,9,12,15,30,120),\sigma$, where $\sigma = 10~\mu\mathrm{Jy/beam}$.), which are identical to the black contours in panel (b). (b) Total intensity radio map of the HH~80-81 region at the VLA C band, where the beam size is  $9''.5 \times 6''.0$. (c) Linearly polarized map towards HH~80 and HH~81, with Stokes I contours. The cyan ellipse at the bottom-left corner of the middle panel shows the beam size. The colour bar is shown on the top of the image.}
    \label{fig:1}
\end{figure*}

%%%%%%%%%% Fig 1 ends %%%%%%%%%%%%%

\section{Observations and Data Reduction} \label{sec:Observations}

We conducted low-frequency radio observations of the massive protostar IRAS~18162$-$2048 (I18162) 
in the C~band using the Karl G. Jansky Very Large Array of the National Radio Astronomy Observatory (Project~ID:~18B-029).  The observations were performed over two separate sessions on 20 and 21 December 2018, with a total on-source time of $\sim$8~hr ($\sim$4~hr per run).  
The phase center was set to $\alpha$(J2000)~$=18^\mathrm{h}19^\mathrm{m}06\fs4$, $\delta$(J2000)~$-20^{\circ}~51'~32'.0$ in the vicinity of HH~80 and HH~81.  For both runs, the phase calibrator J1911$-$2006 was observed, while J2355$+$4950 served as the polarization calibrator.  Flux density and bandpass calibration for each session were performed using~3C286.  Each measurement set contained 32 spectral windows (SPWs), each with 64~channels, 
yielding a total bandwidth of 128~MHz in each spectral window. Prior to calibration and imaging, corrupted data were flagged to mitigate the effects of radio-frequency interference (RFI), non-operational antennas, and time-dependent instrumental issues. Data reduction and imaging were performed using the \texttt{CASA} package \citep{2022PASP..134k4501C}. The absolute polarization angle was determined from observations of 3C286. The antenna-based leakage terms were derived from the polarization calibrator J2355$+$4950, which exhibits stable flux and low intrinsic polarization at the observed frequencies. The polarization calibration was validated by comparing the measured polarization angles and degrees of the flux and phase calibrators with the values listed in the VLA Polarization Database \citep{2017ApJS..230....7P}. This was found to be in excellent agreement. The data from observing sessions on both days were concatenated, and 10 clean SPWs were imaged individually using the \texttt{TCLEAN} task with multiscale deconvolution and \texttt{nterm}~=~2 omitting RFI affected spectral windows. For consistency with earlier linear polarization (LP) measurements of the HH~80-81 jet \citep{2010Sci...330.1209C}, a $uv$~taper of 20~k$\lambda$ was applied. Primary beam correction was performed for all images. The resulting synthesized beam size for all spectral windows were convolved to $9\farcs5 \times 6\farcs0$ with a position angle of $+2^\circ$. The final Stokes images achieved an rms noise level of $10$--$20~\mu\mathrm{Jy~beam^{-1}}$. The LP intensity map was computed from the Stokes~$Q$ and~$U$ images as $\Pi_{L} = \sqrt{Q^2 + U^2}$.

\section{Results} \label{sec:Results}

The total intensity (Stokes~$I$) radio continuum image of I18162 reveals the massive protostar and its bipolar radio jet, featuring multiple emission knots, including the Herbig-Haro objects HH~80 and HH~81.  
This emission is shown in Fig~\ref{fig:1}(b). The brightest continuum emission is centered on the massive protostar I18162, and the source names follow \citet{1993ApJ...416..208M}. The corresponding linearly polarized emission is detected at smaller scales ($\sim$~0.4 pc from the protostar) across multiple frequencies, spanning $4-6$~GHz, in both the jet (S~15) and the counterjet (S~13) (Fig.\ref{fig:1} (a)). The fractional polarization distribution is broadly in agreement with the results previously reported by \citet{2025ApJ...978L..31R}. No significant LP is detected towards I18162 itself, consistent with earlier results \citep{2010Sci...330.1209C,2025ApJ...978L..31R,2025ApJ...988L...9C} obtained at comparable rms and sensitivity. Towards the jet lobes HH~80 and HH~81, we detect no emission in the linearly polarized map which is presented in Fig~\ref{fig:1} (c). The rms noise level in the linearly polarized image is $\sigma \approx 10~\mu\mathrm{Jy~beam^{-1}}$, which allows us to place a $3\sigma$ upper limit of $\sim 30~\mu\mathrm{Jy~beam^{-1}}$ on the linearly polarized intensity in these regions. Thus if linear polarization in HH~80 and HH~81 is present, it is below the detection threshold of our observations.

\section{Discussion} \label{sec:Discussion}

We now address the absence of detectable LP emission toward the outer knots (HH~80 and HH~81) in contrast to the distinct linearly polarized emission observed in the inner jet knots. Several physical mechanisms have been proposed in the literature to explain the presence of linear polarization in protostellar jets. In the following, we examine each of these mechanisms in turn, with the aim of providing a comprehensive interpretation of the conditions prevailing in the outer lobes of the HH~80-81 jet.

\subsection{Acceleration efficiency}

A potential mechanism to consider is synchrotron emission, which produces linearly polarized radiation when relativistic electrons spiral around magnetic field lines \citep{1970ranp.book.....P}. In the case of protostellar jets, synchrotron emission is expected to arise in regions where shocks efficiently accelerate particles to relativistic energies. The degree of linear polarization and the observed spectral index depend on the underlying electron energy distribution and the magnetic field geometry. Measured spectral indices in HH~80 and HH~81 are negative (with $\alpha = -0.3$ to $-0.6$ for HH~80 and $\alpha = -0.1$ to $-0.3$ for HH~81), consistent with non-thermal synchrotron emission \citep{1993ApJ...416..208M,2012ApJ...758L..10M,2018MNRAS.474.3808V}. We examine whether electrons can be efficiently accelerated in both the bow shock and the reverse shock (Mach disk) of the HH~80--81 jet. To determine whether a given shock is adiabatic or radiative, we adopt the criterion proposed by \citet{1989ApJ...337L..37B}, which defines the dimensionless cooling parameter $\chi_\mathrm{s} = d_{cool}/r_{jet}$, which compares the thermal cooling distance ($d_{cool}$) with the jet radius ($r_{jet}$) at the shock location. Here, $d_{cool}$ represents the distance behind a steady-state radiative shock at which gas entering the shock at velocity $v_{s}$ cools to $\sim 10^{4}~\mathrm{K}$. When $\chi_\mathrm{s} \gg 1$, the shock-heated gas does not have time to cool before leaving the working surface, and the shock is effectively adiabatic. Conversely, for a fully radiative shock ($\chi_\mathrm{s} \ll 1$), the post-shock gas loses its thermal energy within a relatively short distance downstream of the shock.  For the jet radius, we measure a value of $r_{\mathrm{jet}} \approx 7~\times~10^{16}~\mathrm{cm}$ and $3~\times~10^{16}$~cm, corresponding to the deconvolved semi-minor axis of a gaussian fit to HH~80 and HH~81 which are consistent with the linear sizes obtained using X-ray observations ( $r_{\mathrm{jet}} \approx 7.5~\times~10^{16}~\mathrm{cm}$ and $5~\times~10^{16}$~cm for HH~80 and HH~81 respectively) \citep{2004ApJ...605..259P} and radio observations by \citet{2019MNRAS.482.4687R}. The cooling distance downstream of a shock with velocity $v_{\mathrm{s}}$ propagating into a medium with density $n$ can be estimated as:

\begin{equation}\label{eqn:3}
\left ( \frac{d_{\mathrm{cool}}}{\mathrm{cm}}  \right) \;\approx\;
1.8 \times 10^{14}\;
\left( \frac{v_s}{100\ \mathrm{km\ s^{-1}}} \right)^{4.67}
\left( \frac{100\ \mathrm{cm^{-3}}}{n} \right)
\end{equation}
for $v_s > 60$~km s$^{-1}$ \citep{1987ApJ...316..323H} and,

\begin{equation}\label{eqn:4}
\left ( \frac{d_{\mathrm{cool}}}{\mathrm{cm}}  \right) \;\approx\;
2.24 \times 10^{14}\;
\left( \frac{v_s}{100\ \mathrm{km\ s^{-1}}} \right)^{4.5}
\left( \frac{100\ \mathrm{cm^{-3}}}{n} \right)
\end{equation}
for $v_s > 400$~km s$^{-1}$ \citep{2002ApJ...576L.149R}.

Following \citet{1990ApJ...360..370B}, the velocity of the bow shock, $v_{\mathrm{bs}}$, can be estimated by equating the momentum flux of the jet beam (the collimated supersonic flow) at the working surface with the momentum flux of the ambient medium at the bow shock:

\begin{equation}\label{eqn:5}
v_{\rm bs} \;=\; \frac{v_{\rm jet}}{1 + \eta^{-1/2}},~\eta \;=\; \frac{n_{\rm jet}}{n_{\rm amb}}
\
\end{equation}

For the ambient density, we adopt an average value as measured by \citet{1986ApJ...307..787T}, 
$n_{\mathrm{amb}} = 5~\times~10^{3}~\mathrm{cm^{-3}}$. We take the jet velocity to be $v_{\mathrm{jet}} = 1000~\mathrm{km~s^{-1}}$ \citep{1995ApJ...449..184M}. The jet density is given as $1000$~cm$^{-3}$ for HH~81 and $200$~cm$^{-3}$ for HH~80 \citep{1998AJ....116.1940H,1998ApJ...501..687G,2023JApA...44...57M}. Substituting these values into Eqn.~(\ref{eqn:5}), we obtain a bow-shock velocity of  $v_{\mathrm{bs}} \approx 166~\mathrm{km~s^{-1}}$ and $309~\mathrm{km~s^{-1}}$ for HH~80 and HH~81, respectively.  Using this velocity in Eqn.~(\ref{eqn:3}), we find a ratio $\chi_\mathrm{s} \approx $ $0.023$ for HH~80 and $0.906$ for HH~81,  which is below unity.  This indicates that the bow shock is radiative. In such shocks, Alfvén waves are strongly damped, and diffusive shock acceleration (DSA) is expected to be inefficient.

\begin{figure}
    \centering
    \includegraphics[width=\linewidth]{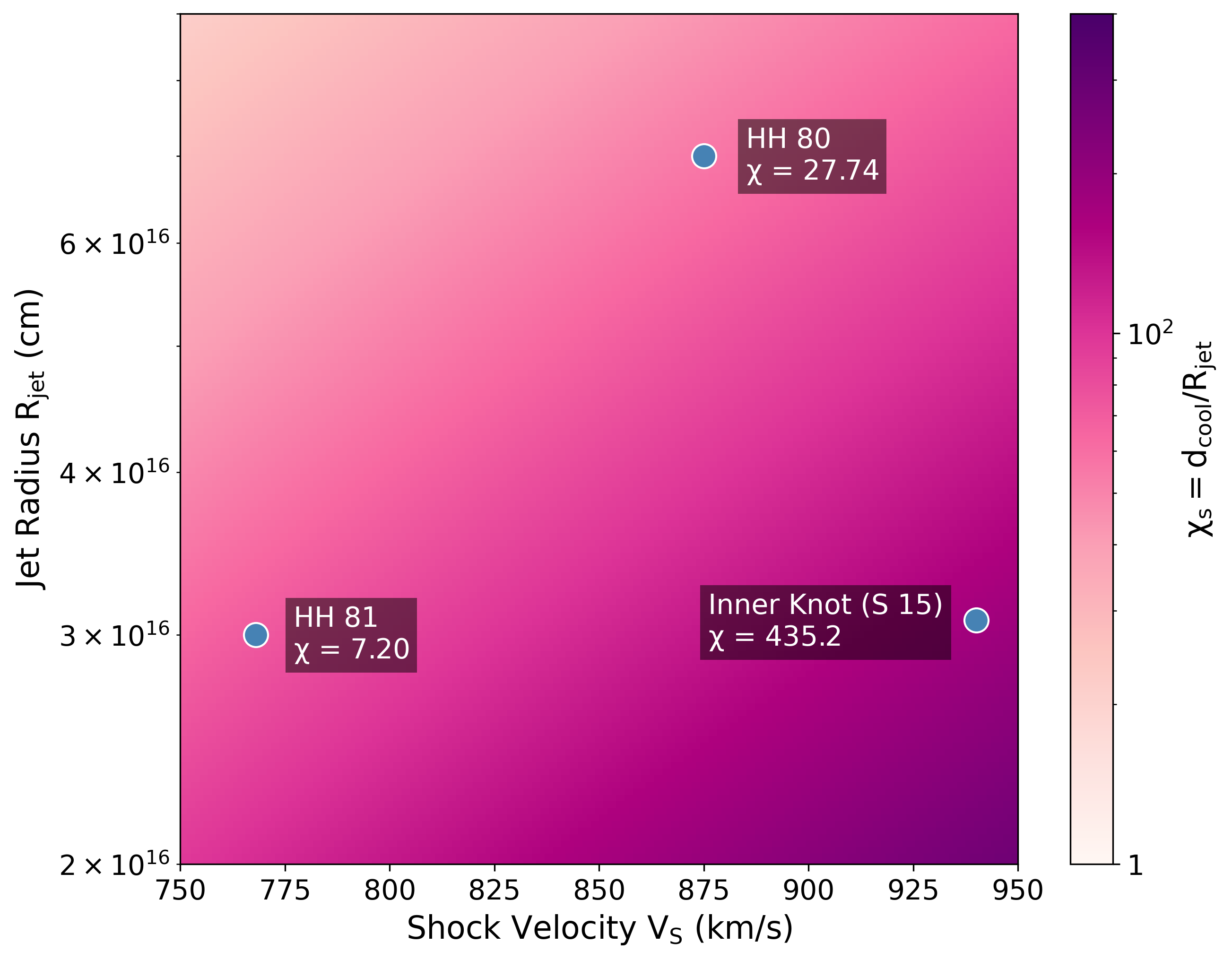}
    \caption{A plot of the shock velocity ($v_{s}$) as a function of the jet radius ($r_{\rm jet}$), with $\chi_\mathrm{s}$ values shown using a colour scale for comparison between the inner knots and HH~80-81. The $\chi_\mathrm{s}$ values derived for each knot are indicated by grey points, and the corresponding colour bar is displayed on the right-hand side of the plot.}
    \label{fig:2}
\end{figure}

To evaluate whether particle acceleration can be efficient in the reverse shock,  we first estimate the shock velocity, $v_{\mathrm{rs}}$. Under the assumption of pressure equilibrium between the shocked regions, the reverse-shock velocity can be expressed as \citep{1990ApJ...360..370B}
\begin{equation}\label{eqn:6}
v_{\mathrm{rs}} = v_{\mathrm{jet}} - \frac{3}{4} v_{\mathrm{bs}} 
\end{equation}
Using the values adopted in the previous section for $v_{\mathrm{bs}}$ and $v_{\mathrm{jet}}$, we obtain $v_{\mathrm{rs}} \approx 875~\mathrm{km~s^{-1}}$ for HH~80 and $768~\mathrm{km~s^{-1}}$ for HH~81. The corresponding thermal cooling distance, calculated from Eqn.~(\ref{eqn:4}) with $n = n_{\mathrm{jet}}$, yields $d_{cool} = 1.9~\times~10^{18}$~cm and $2.1~\times~10^{17}$~cm. The ratio $d_{cool}/r_{jet}$ for HH~80 and HH~81 knots are $\chi_\mathrm{s} \approx 27.7$ and $7.2$ respectively. This ratio, being larger than unity, implies that the shock can be adiabatic in both cases of HH~80 and HH~81, as summarized in Table~\ref{Tab:1}. In such conditions, diffusive shock acceleration can operate efficiently, potentially accelerating particles to relativistic energies. Towards the inner knots (S~15) the cooling parameter reaches a value of  $\chi_\mathrm{s} \approx 472$ \citep{2017ApJ...851...16R}, which is significantly higher than those obtained for HH~80 and HH~81 knots. The larger value in the inner knot(s) suggests a greater efficiency in the acceleration process, consistent with the enhanced polarized emission observed towards the inner knots. This is illustrated in the plot shown in Fig.~\ref{fig:2}.

%%%%%%%%%% Fig 1 begins %%%%%%%%%%%%%

%%%%%%%%%% Fig 1 ends %%%%%%%%%%%%%

We now compare the energy requirements for electron acceleration. The acceleration timescale in the reverse shock is given by  \citep{1983RPPh...46..973D}

\begin{equation} \label{eqn:7}
t_{\rm acc} = \epsilon \frac{E}{e B c}, \qquad 
\epsilon = \frac{20}{3}\left(\frac{c}{v_{\rm rs}}\right)^{2}
\end{equation}
where, $B$ is the magnetic field strength, $E$ is the particle energy, and $v_{\rm rs}$ is the reverse-shock velocity given by Eqn.~(\ref{eqn:6}). If synchrotron losses dominate, the maximum attainable electron energy $E_{\max}$ is determined by the condition $t_{\rm acc} = t_{\rm synch}$, where the synchrotron loss timescale given by \citet{1983RPPh...46..973D} is  

\begin{equation}\label{eqn:8}
t_{\rm synch} \simeq 4\times 10^{11}
\left(\frac{B}{\rm mG}\right)^{-2}
\left(\frac{E}{\rm GeV}\right)^{-1} \, {\rm s}
\end{equation}

Equating the two timescales yields  

\begin{equation}\label{eqn:9}
E_{\max} \simeq 2.4 \times 10^{3}\, 
\left(\frac{v_{\rm rs}}{10^{8}\,{\rm cm\,s^{-1}}}\right) 
\left(\frac{B}{\rm mG}\right)^{-1/2} \, {\rm GeV}
\end{equation}

For HH~80 and HH~81, we consider a magnetic field strength of $0.13$~mG and $0.16$~mG \citep{2018MNRAS.474.3808V,2023JApA...44...57M}, and assume Bohm diffusion and we obtain $\epsilon \simeq 6.3\times10^{6}$ and $20.5\times10^{6}$, corresponding to maximum electron energies of $E_{\max} \approx 6.6~{\rm TeV}$ and $5.8~{\rm TeV}$, respectively. In contrast, for the inner knots we find 
$E_{\max} \approx 2~{\rm TeV}$ \citep{2017ApJ...851...16R}. This comparison indicates that significantly lower energy is required to accelerate electrons in the inner knots relative to outer lobes, making the production of synchrotron emission more favourable in the inner jet regions.

%%%%%%%%% Beg TABLE 1 %%%%%%%%%%%%%%%%%%%%
\begin{table*}[ht]
    \caption{Details of the shock types, physical parameters and estimated properties for the jet knots HH~80 and HH~81.}
    \hspace*{-2cm}
    \label{Tab:1}
    \centering
    \begin{tabular}{ccccccc}
        \hline
        \multirow{2}{*} {Source} & Jet radius & Number density & Shock type & Shock velocity  & $d_{cool}$\tablenotemark{a,b} & $\chi_s$\tablenotemark{c} \\
 & (cm) & (cm$^{-3}$)& & (km/s) & (cm) & \\
\hline
HH~80 & $(7\pm0.3) \times 10^{16}$ & $200\pm10$ & Bow Shock & $166.7\pm17.4$ & $(3.8\pm1.9) \times 10^{13}$ & $0.02\pm0.01$ \\
      & & & Reverse Shock & $875.5\pm87.6$ & $(1.9\pm0.8) \times 10^{18}$ & $27.7\pm12.5$ \\
HH~81 & $(3\pm0.1) \times 10^{16}$ & $1000\pm50$ & Bow Shock & $309.0\pm31.8$ & $(6.9\pm3.3) \times 10^{14}$ & $0.90\pm0.09$ \\
      & & & Reverse Shock & $768.2\pm77.0$ & $(2.1\pm0.1) \times 10^{17}$ & $7.2\pm3.2$ \\
Inner Knot (S~15)\tablenotemark{d} & $(3\pm0.1) \times 10^{16}$ & $40$ & Bow Shock & $80$ & $1.2 \times 10^{12}$ & $4\times10^{-5}$ \\
      & & & Reverse Shock & $940$ & $1.3 \times 10^{19}$ & $432$ \\
\hline
    \end{tabular}

\smallskip
\tablenotetext{a}{Cooling length $
\left ( {d_{\mathrm{cool}}}/{\mathrm{cm}}  \right) \;\approx\;
1.8 \times 10^{14}\;
\left( {v_s}/{100\ \mathrm{km\ s^{-1}}} \right)^{4.67}
\left( {100\ \mathrm{cm^{-3}}}/{n} \right)$ for $v_{s} > 60$ km~s$^{-1}$} 
\tablenotetext{b} {Cooling length $
\left ( {d_{\mathrm{cool}}}/{\mathrm{cm}}  \right) \;\approx\;
2.24 \times 10^{14}\;
\left( {v_s}/{100\ \mathrm{km\ s^{-1}}} \right)^{4.5}
\left( {100\ \mathrm{cm^{-3}}}/{n} \right)$ for $v_{s} > 400$ km~s$^{-1}$}
\tablenotetext{c}{Cooling parameter $(\chi_s) = d_{cool}/r_{jet}$.}
\tablenotetext{d}{Parameters for the Inner Knot (S~15) are adopted from \citet{2017ApJ...851...16R}.}
\end{table*}

%%%%%%%%% End TABLE 1 %%%%%%%%%%%%%%%%%%%%

\subsection{Depolarization}

In addition to shock acceleration, depolarization can be an important factor that can significantly influence the observed level of polarization \citep{1970ranp.book.....P}. Depolarization may arise from several mechanisms, including Faraday rotation within a magnetized medium, where variations in the electron density or magnetic field along the line of sight cause different parts of the emitting region to experience different rotation angles, thereby reducing the net polarized signal \citep{1966MNRAS.133...67B,1977A&A....61..291J,1998MNRAS.299..189S,2021ApJ...923L...5P}. Beam depolarization can also occur when unresolved substructures within the telescope beam possess different polarization orientations that average out in the observed emission \citep{1966MNRAS.133...67B}. Furthermore, turbulent magnetic fields in the shocked region can lead to spatial variations in the polarization angle, producing additional cancellation \citep{1998MNRAS.299..189S,2002A&A...388.1106B}. These effects can diminish the measured polarization fraction, even in regions where synchrotron emission is intrinsically polarized, and thus must be taken into account when interpreting the polarized emission in HH~80-81 and in the inner knots.

To evaluate the extent of this depolarization, it is useful to compare the theoretical intrinsic polarization. For optically thin synchrotron emission, the maximum intrinsic fractional linear polarization depends on the electron energy index \(p\) (with \(N(E)\!\propto\!E^{-p}\)) or, equivalently, on the spectral index \(\alpha\) (\(S_\nu\!\propto\!\nu^{-\alpha}\), with \(p=2\alpha+1\)). The intrinsic value is
\begin{equation}\label{eqn:10}
\Pi_0=\frac{p+1}{p+\tfrac{7}{3}}
=\frac{3\alpha+3}{3\alpha+5}
\end{equation}

which yields \(\Pi_{\rm 0}\!\approx\!0.63\)–\(0.66\) for \(\alpha\!\sim\!0.5\)–\(1\) \citep{1970ranp.book.....P,1979rpa..book.....R,1995ApJ...449..184M,2018MNRAS.474.3808V,2023JApA...44...57M}. (For self-absorbed/optically thick synchrotron, \(\Pi_{\rm 0}\) is much smaller, \(\sim\!10\%\), with the fractional polarization of the inner knots are $15-30\%$ \citep{2010Sci...330.1209C}. 

Even if we adopt the theoretical polarization fraction as an upper limit, the deficit can be quantified in terms of fluctuations in the rotation measure (RM). In the case of an external Faraday screen with a Gaussian distribution of RM values \citep{1966MNRAS.133...67B}, the depolarization factor is
\begin{equation}\label{eqn:11}
\frac{\Pi_{\max}}{\Pi_{0}} = \exp\!\left(-2 \,\sigma_{\mathrm{RM}}^{2}\,\lambda^{4}\right),
\end{equation}

where \(\Pi_{\max}\) is the observed maximum fractional polarization, \(\Pi_{0}\) is the intrinsic value, and \(\lambda\) is the observing wavelength. Solving for the RM dispersion gives
\begin{equation}\label{eq:12}
\sigma_{\mathrm{RM}} \;=\; \sqrt{ \frac{-\ln\!\left(\Pi_{\max}/\Pi_{0}\right)}{2\,\lambda^{4}} } .
\end{equation}

Here, the observing wavelength is taken to be $\lambda= 6~\mathrm{cm}~( \nu =5$~GHz) and $\Pi_{\mathrm{max}}$ = P$_\mathrm{{min}}$/I, where P$_\mathrm{{min}}$ is the upper limit of polarization and I is the Stokes~I flux density. We adopt an upper limit of $30~\mu$Jy and the flux densities for HH~80 and HH~81 are 1.14~mJy and 1.68~mJy respectively. Substituting these, we obtain $\Pi_{\mathrm{max}}$ = $0.026$ and $0.017$ respectively. Substituting these into Eqn.~(\ref{eq:12}), we yield $\sigma_\mathrm{{RM}} \sim 350 $~rad/m$^2$ - $400$~rad/m$^2$. For comparison, the rotation measure values reported by \citet{2025ApJ...978L..31R} are in the range 200--500~rad~m\(^{-2}\). It is noteworthy that the derived values of $\sigma_{\mathrm{RM}}$ are comparable to, or in some cases exceed, the measured rotation measure (RM) values themselves. This indicates that the magnitude of the RM fluctuations is at least as large as the mean RM along the line of sight. Such a condition implies that the dispersion in Faraday rotation is not a minor perturbation but instead dominates over the underlying RM signal. Consequently, the observed polarization is subject to strong depolarization effects, with small-scale fluctuations in the magneto-ionic medium effectively washing out a substantial fraction of the intrinsic synchrotron polarization.

This highlights the critical role of depolarization in shaping the observed polarization properties of HH~80-81 and the inner knots: even if synchrotron emission is intrinsically highly polarized, Faraday fluctuations and beam averaging can suppress the measurable signal. Therefore, direct measurements of polarized emission from jet knots are indispensable for disentangling these effects. Such observations not only probe the efficiency of shock compression and particle acceleration, but also provide unique diagnostics of the magnetic-field topology, turbulence, and ionized environments along the line of sight. High-resolution, broadband polarimetric studies are especially important, as they can mitigate beam depolarization, quantify rotation measure dispersions, and recover the intrinsic polarization fraction. 

Future radio facilities such as the \textit{Square Kilometre Array Observatory} (SKAO) will be transformative in this respect. With its unprecedented sensitivity, wide frequency coverage, and sub-arcsecond angular resolution, SKAO will enable the detection and mapping of faint polarized emission from protostellar jets, including HH~80-81. This will allow systematic characterization of Faraday depolarization, RM variations, and turbulent magnetic fields, thereby offering a new window into the magnetized environments of young stellar objects and their powerful jets.

\section{Conclusion} \label{sec:Conclusion}

We have studied the linearly polarized radio emission from HH~80-81 using VLA observations in 4-6~GHz. No polarized emission was detected at the knot positions; adopting a conservative upper limit on the polarized intensity of $\rm P_{\rm min}=30~\mu\mathrm{Jy}$ yields maximum fractional polarizations $\Pi_{\max}\approx0.02$ and $0.01$ for HH~80 and HH~81 respectively. To understand this non-detection we examined both intrinsic and observational effects. From synchrotron theory the expected intrinsic fractional polarization for optically thin emission is $\Pi_{0}\sim0.63$ - $0.66$ for plausible spectral indices, implying a large deficit between the theoretical and observed limits. 

We evaluated shock and cooling conditions through the parameter $\chi_\mathrm{s}=d_{\rm cool}/r_{\rm jet}$ and found values for HH~80-81 that are much smaller than those measured in the inner knots. This suggests that the reverse shocks in HH~80-81 are less efficient in diffusive shock acceleration than the inner knots, and hence intrinsically weaker polarized synchrotron emission. In addition, our derived RM dispersions, $\sigma_{\rm RM}\sim 350$--$400$ rad/m$^2$, are comparable to or exceed reported RM values, indicating substantial small-scale Faraday fluctuations. Such fluctuations, together with beam depolarization from unresolved substructure and turbulent magnetic fields, can strongly suppress observable linear polarization even when intrinsic polarization is high. 

We therefore conclude that the lack of detected polarization in HH~80-81 results from a combination of (i) intrinsically less efficient particle acceleration compared with the inner knots and (ii) significant observational depolarization (Faraday and beam averaging). Future deep, high-resolution, broadband polarimetric observations---particularly with next-generation facilities such as the SKAO---are required to disentangle intrinsic emission from depolarization effects and to robustly map the magnetic and particle acceleration properties along these protostellar jets.

\section{Acknowledgments} \label{sec:Acknowledgments}

We thank the referee for carefully perusing the manuscript and giving inputs that have helped improve the paper. We thank the staff of Karl G. Jansky Very Large Array (VLA), who made the radio observations possible. The National Radio Astronomy Observatory is a facility of the National Science Foundation operated under a cooperative agreement by Associated Universities, Inc. A.G.-C. acknowledges the Indian Institute of Science (IISc), Bengaluru, where partial data analysis was carried out. N.R. acknowledges support from the United States-India Educational Foundation through the Fulbright Program. C.C.-G. acknowledges support from UNAM DGAPA-PAPIIT grant IG101224. A.P. acknowledges support from UNAM DGAPA-PAPIIT grant IA100425.

\bibliography{sample701}{}
\bibliographystyle{aasjournal}

\end{document}